\documentstyle[pra,aps,multicol,epsfig]{revtex}
\begin{document}

\title{Macroscopic quantum bound states of Bose Einstein
condensates in optical lattices }

\author{Mario Salerno}

\address{Dipartimento di Fisica "E.R. Caianiello"
         and Istituto Nazionale di Fisica della Materia (INFM), \\
         Universit\'a di Salerno, I-84081 Baronissi (SA),
         Italy}

\date{\today}
\maketitle

\begin{abstract}

We discuss localized ground states of the periodic
Gross-Pitaevskii equation in the framework of a quantum linear
Schr\"odinger equation with effective potential determined in
self-consistent manner. We show that depending on the interaction
among the atoms being attractive or repulsive, bound states of the
linear self consistent problem  are formed in the forbidden zones
of the linear spectrum below or above the energy bands. These
eigenstates are shown to be exact solitons of the GPE equation.
The implications of this bound state interpretation on the
existence of a delocalization transition for multidimensional
solitons is briefly discussed.

\end{abstract}


\begin{multicols}{2}

One interesting phenomenon occurring in periodic nonlinear systems
is the possibility to stabilize localized excitations as a result
of the interplay between periodicity and nonlinearity. An example
of this is provided by the nonlinear Schr\"odinger equation (NLS)
with periodic potential. It is well known that the defocusing NLS
does not admit bright soliton solutions, these being unstable
against background decay \cite{scott}. The presence of a periodic
potential, however, allows to stabilize bright solitons against
decay, a phenomenon which is presently investigated in connection
with Bose Einstein condensates (BEC) in optical lattices (OL). The
possibility to form bright solitons in repulsive BEC with OL was
analytically and numerically demonstrated, both for a discrete
version of the NLS describing BEC arrays in the tight-binding
approximation \cite{smerzi} and for the Gross-Pitaevskii equation
(GPE) describing the properties of a continuous BEC  in the mean
field approximation \cite{potting,ks02,alfimov}. The mechanism
underlying soliton formation in periodic structures was identified
to be the modulational instability of the Bloch states at the
edges of the Brillouin zone \cite{ks02}. These localized
excitations correspond to states with energies inside the gaps of
the underlying linear band structure (in nonlinear optics they are
called gap solitons) and with an  effective mass which depends on
the sign of the interaction (for repulsive interactions, bright
solitons have negative effective mass, this explaining their
existence in BEC with OL \cite{ks02,steel}). The usage of linear
concepts such as Bloch states, effective mass, etc.
\cite{ks02,steel,pethick}, makes natural to ask whether nonlinear
states could be interpreted  in a pure linear (quantum mechanical)
context.

The aim of the present paper is to address this problem by showing
that soliton solutions of the periodic nonlinear Schr\"odinger
equations correspond to bound states of the linear Schrodinger
equation with an effective potential which can be determined in
self-consistent (SC) manner.
%This approach
%appears to be the Landau-Peaker
%theory of the small polaron for nonlinear Schr\"odinger-like
%equations.
This problem will be discussed on the  physical example of a Bose
Einstein condensate in an optical lattice (OL) described, in mean
field approximation, by the following normalized Gross-Pitaevskii
equation
\begin{equation}
i\psi_t=\left[-\nabla ^{2}+ U_{ol}({\bf{x}}) +
\chi|\psi|^{2}\right] \psi \label{gpe}
\end{equation}
where $\chi$ is the nonlinear parameter, $\bf x$ denotes three
dimensional coordinates and $U(\bf x)$ is a periodic potential
representing the OL. To discuss bound state features of solitons
we restrict to the one dimensional case (the approach however is
of general validity and can be applied to NLS type equations in
arbitrary dimensions). At the end of the paper we will briefly
discuss the implications of the bound state interpretation of
localized solutions on the soliton delocalization transition
observed in higher dimensions \cite{flach}. We remark that the
properties of solitons of the GPE in optical lattices were studied
in \cite{alfimov} in terms of orbits of a chaotic system.
Self-consistent approaches were also used as numerical tools to study
discrete breathers of the discrete NLS \cite{panos} and the
stability of gap solitons \cite{markus}. The physical implications
and the full potentiality of the SC approach, however, have not
been investigated.

Our analysis is based on the simple observation that the
stationary localized ground states $\psi_s(x,t) = \psi(x)
\exp(-\mu t)$ of the GPE (and more generally of any nonlinear
Schr\"odinger-like equation) can be obtained by solving in a
self-consistent manner the following  linear Schr\"odinger problem
\begin{equation}
\left[-\nabla ^{2}+ \hat V_{eff}(x) \right] \psi= E \psi
\label{schro}
\end{equation}
with the effective potential
\begin{equation}
\hat V_{eff}= \hat U_{ol}(x)+ \hat U_s (x) = A \cos(2 x) + \chi
|\hat \psi_s (x)|^2.  \label{Veff}
\end{equation}
Here $\hat U_{ol} \equiv A \cos(2 x) $ is the OL and $\hat U_s$ is
the potential associated with a given eigenstate of the quantum
problem (\ref{schro}). For a self-consistent solution, one starts
with a trial wavefunction for $\psi_s$ (typically a gaussian
waveform), calculates the effective potential and solves the
corresponding eigenvalue problem (\ref{schro}). Then, one selects
a given eigenstate (for example the ground state but not
necessarily) as new trial function and iterates the procedure
until convergence is reached.
\begin{figure}\centerline{
\includegraphics[width=3.8cm,height=3.3cm,angle=0,clip]{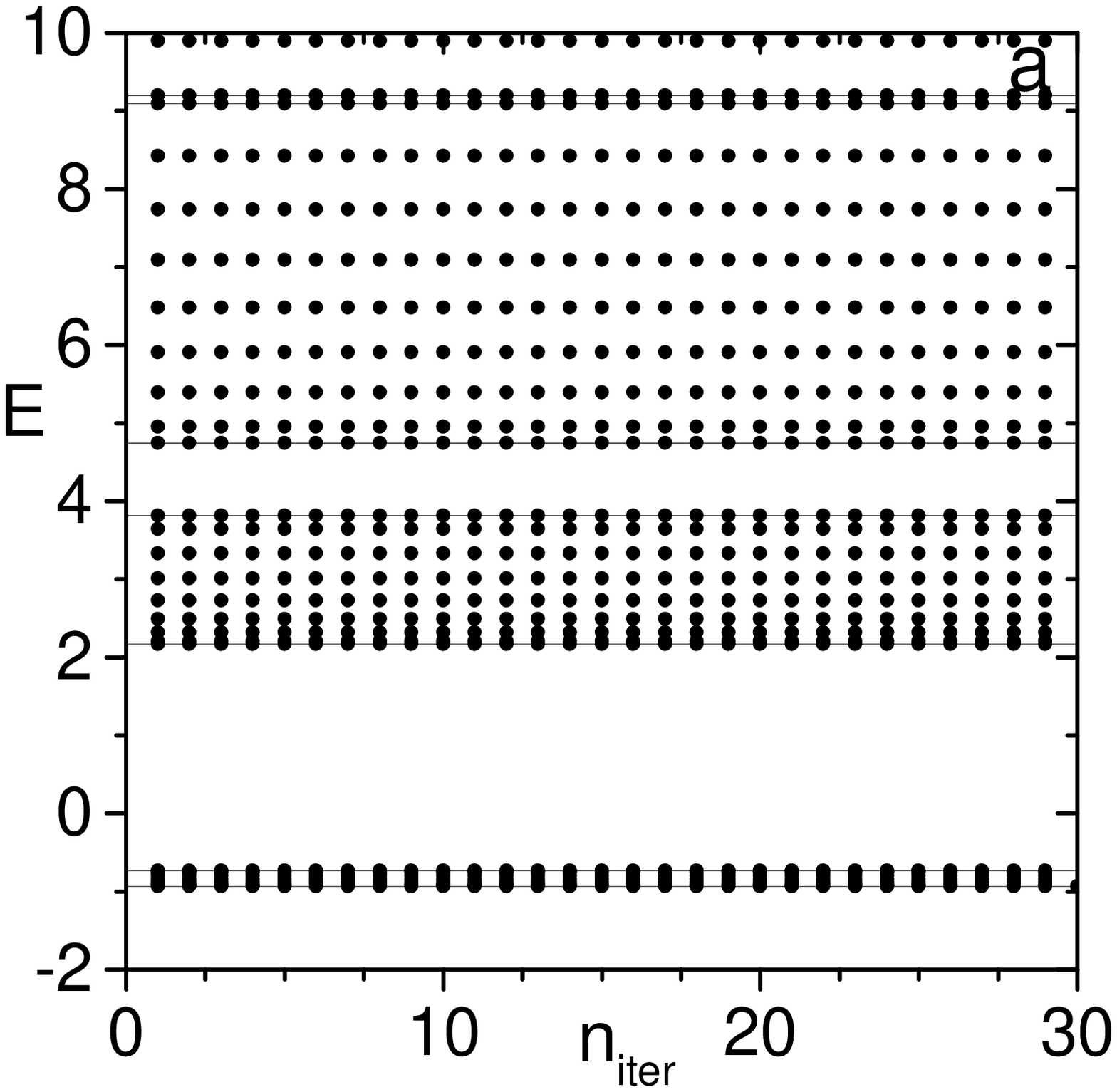}
\includegraphics[width=3.8cm,height=3.3cm,angle=0,clip]{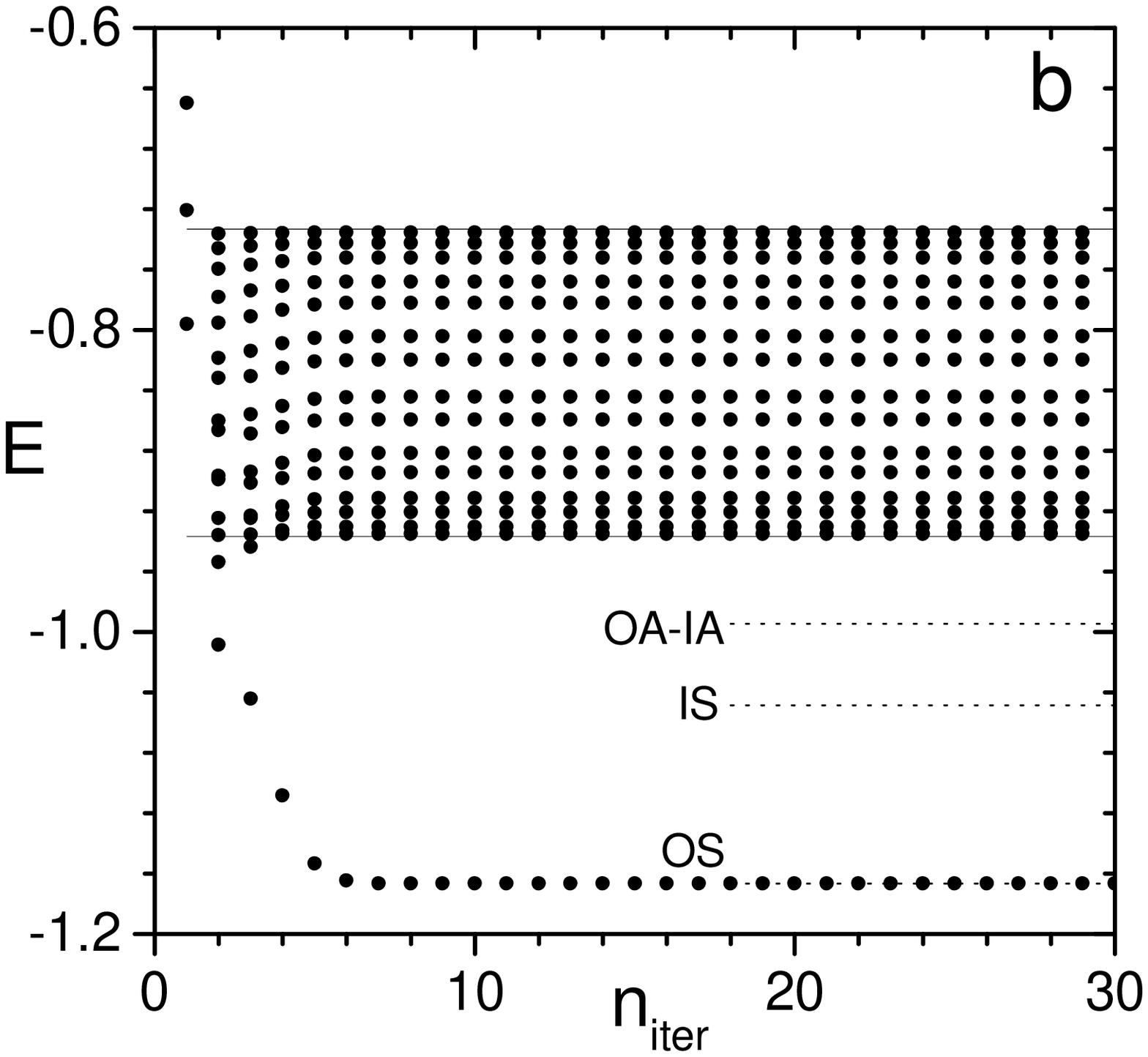}}
\centerline{
\includegraphics[width=3.8cm,height=3.3cm,angle=0,clip]{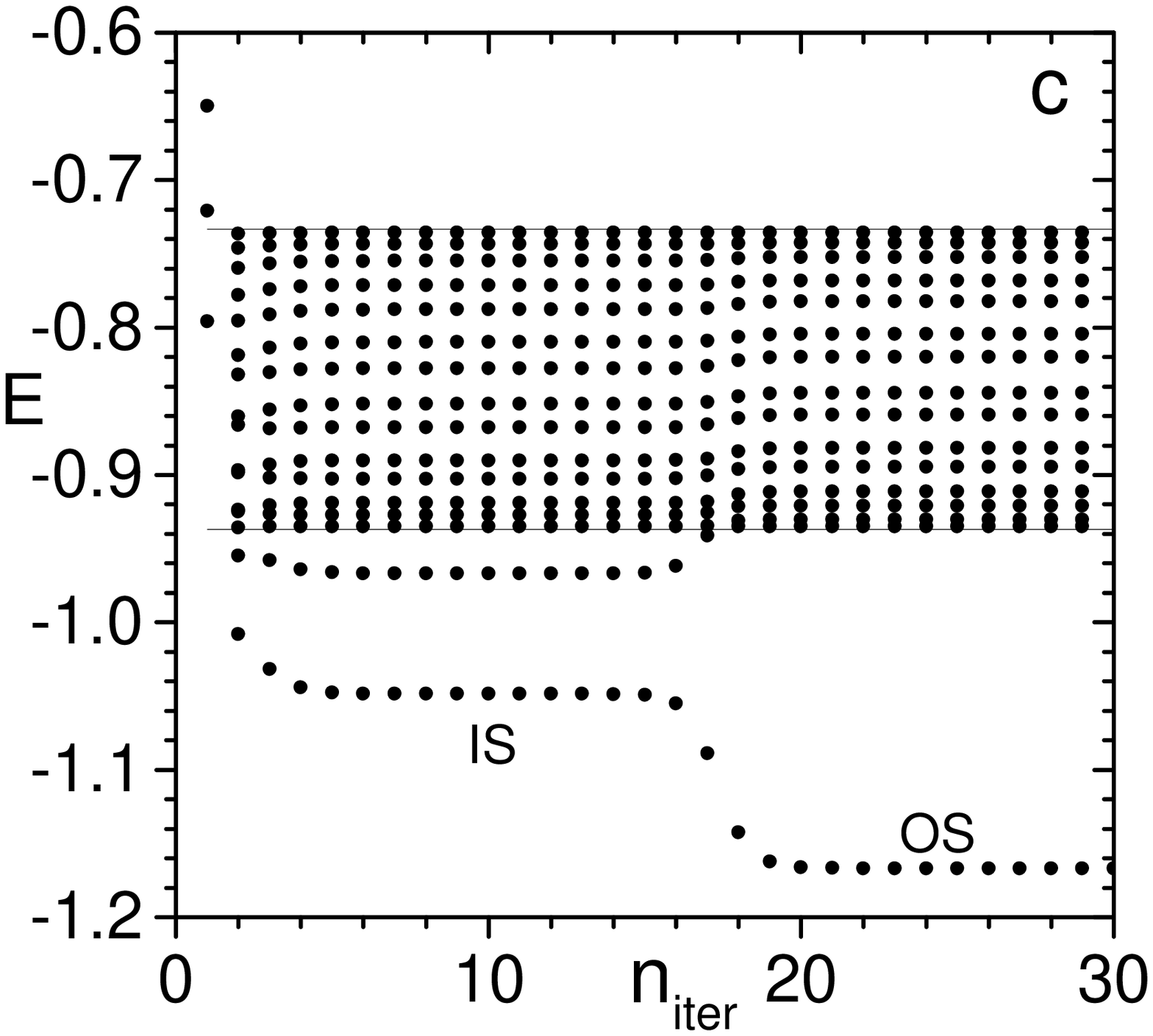}
\includegraphics[width=3.8cm,height=3.3cm,angle=0,clip]{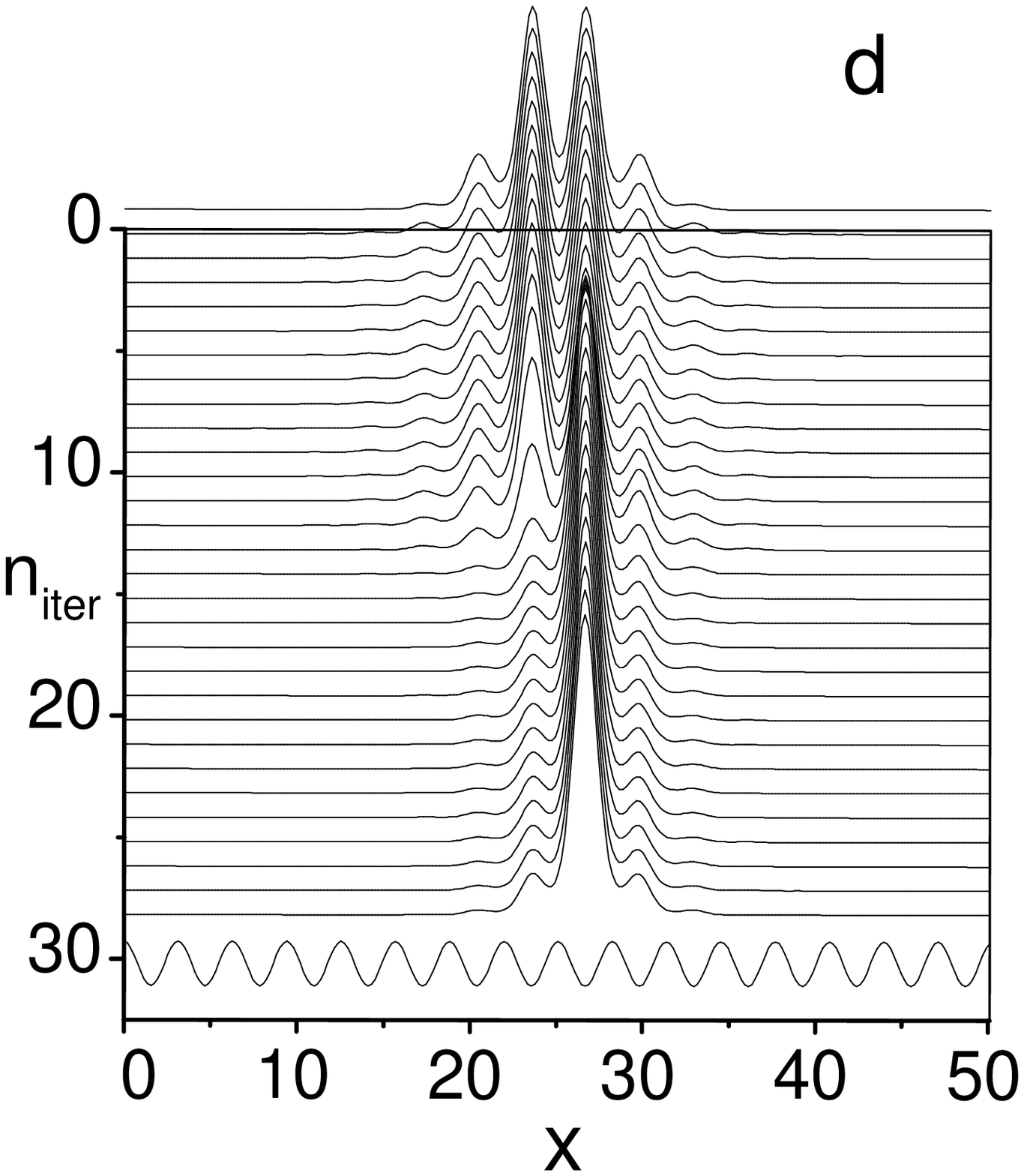}}
\caption{ Panel {\bf (a)} Energy spectrum for the effective
potential (\ref{Veff}) with $A=3$ and $\chi=0$ (Mathieu equation).
Full lines represent exact values of the band edges of the Mathieu
equation while dots are the eigenvalues obtained with the above
procedure on a lattice of length $L=40 \pi$, with $N=512$ points.
Panel {\bf (b)} Lowest energy band for the effective potential in
Eq. (\ref{Veff}) with $\psi_s$ taken as the ground state of the
system and for $\chi=-1$ (attractive case). Parameters are fixed
as in panel (a). Panel {\bf (c)} The same as in panel (b) but for
$A=-3$. Panel {\bf (d)} Transition from the metastable IS mode to
the OS ground state corresponding to the lower level of panel (c).
The optical lattice (scaled by a factor 3) is reported as an help
to locate the symmetry center of the solutions. Parameters are
fixed as in panel (c).} \label{fig1}
\end{figure}

The problem is thus reduced to the diagonalization of the quantum
Hamiltonian

\begin{equation}
\hat H = \hat H_0 + \hat V_{eff} (x) \label{hamilt}
\end{equation}
with $\hat H_0\equiv -\nabla ^{2}$ the kinetic energy operator.
This can be effectively done by adopting a discrete coordinate
space representation $\{x_n=n a\}$, $n=1,...,N$, with $a=L/N$ the
discretization constant, $L$ the size of the system and N the
total number of points. A basis for this space is simply a basis
of $R^N$, i.e. the set of N-component vectors of the type $|n
\rangle=(0,...0,1,0,...,0)$, with the $1$ in the position $n$. The
effective potential is obviously diagonal in this representation
i.e. $\langle n|\hat V_{eff}|n'\rangle = V_{eff} (n a)
\delta_{n,n'}$, while  $\hat H_0$ is diagonal in the reciprocal
representation $|k_n \rangle $, ($k_n= 2\pi/L n $, the two
representations being related by the Fourier transform (unitary
transformation). The matrix elements of the Hamiltonian $\hat H$
can then be constructed as \begin{equation} \langle n|\hat
H|n'\rangle \equiv H_{n,n'}=\langle n| \hat F^{-1} \hat H_0 \hat F
|n'\rangle + V_{eff} (n a) \delta_{n,n'}
\end{equation}
where $\hat F |n \rangle$ denotes the Fourier transform of the
vector $|n \rangle$.
\begin{figure}\centerline{
\includegraphics[width=3.8cm,height=3.cm,angle=0,clip]{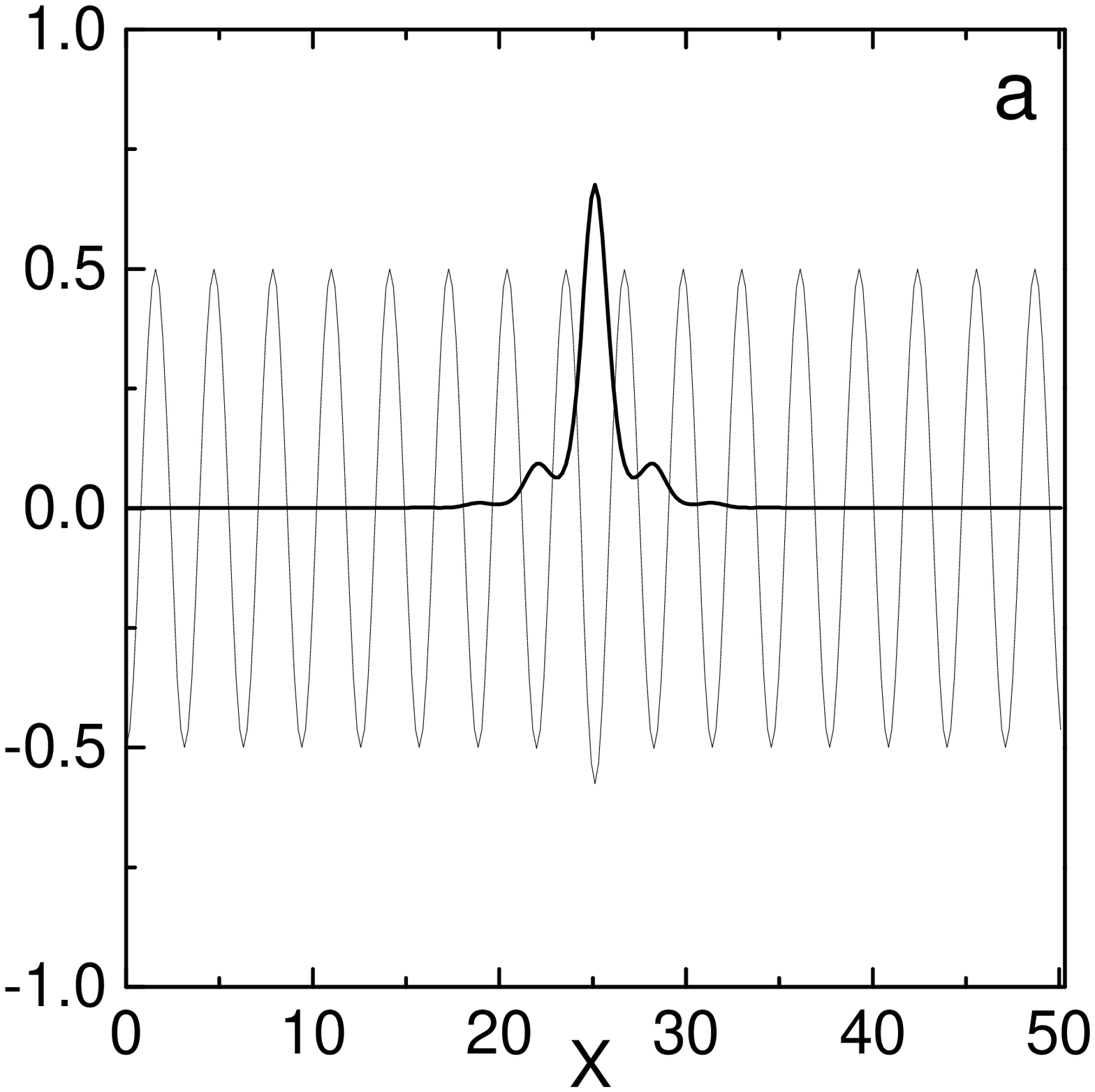}
\includegraphics[width=3.8cm,height=3.cm,angle=0,clip]{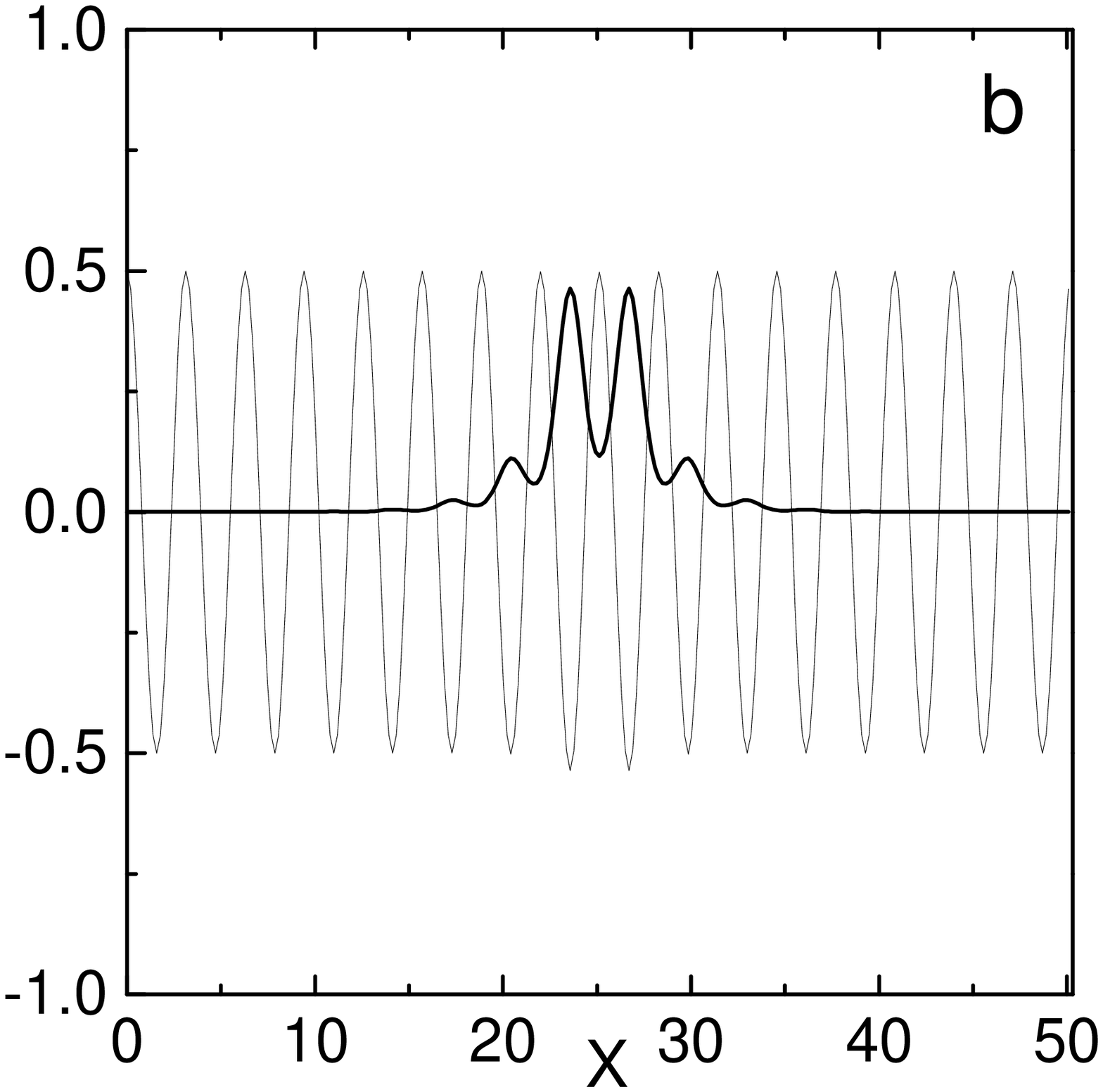}}
\centerline{
\includegraphics[width=3.8cm,height=3.cm,angle=0,clip]{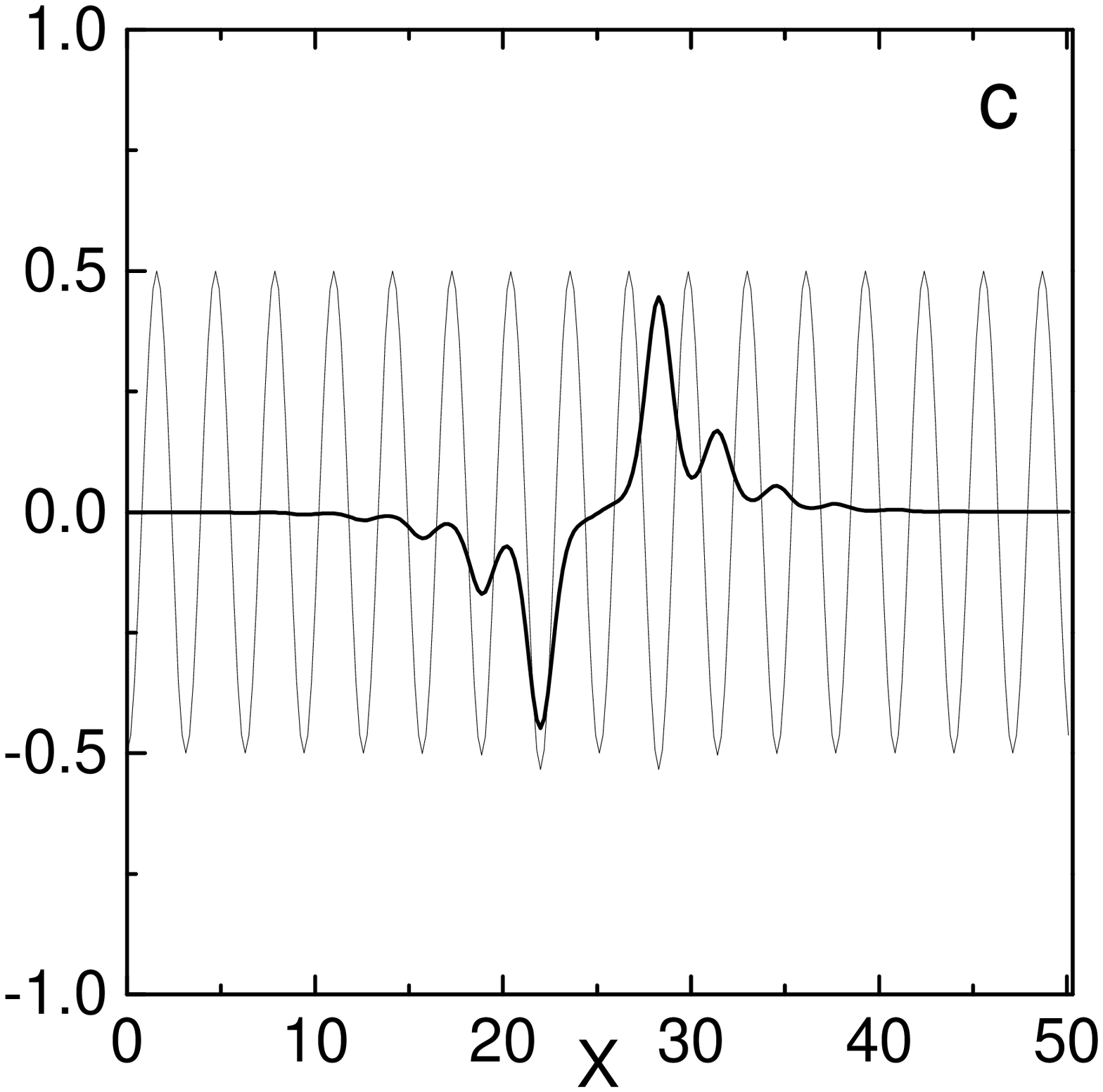}
\includegraphics[width=3.8cm,height=3.cm,angle=0,clip]{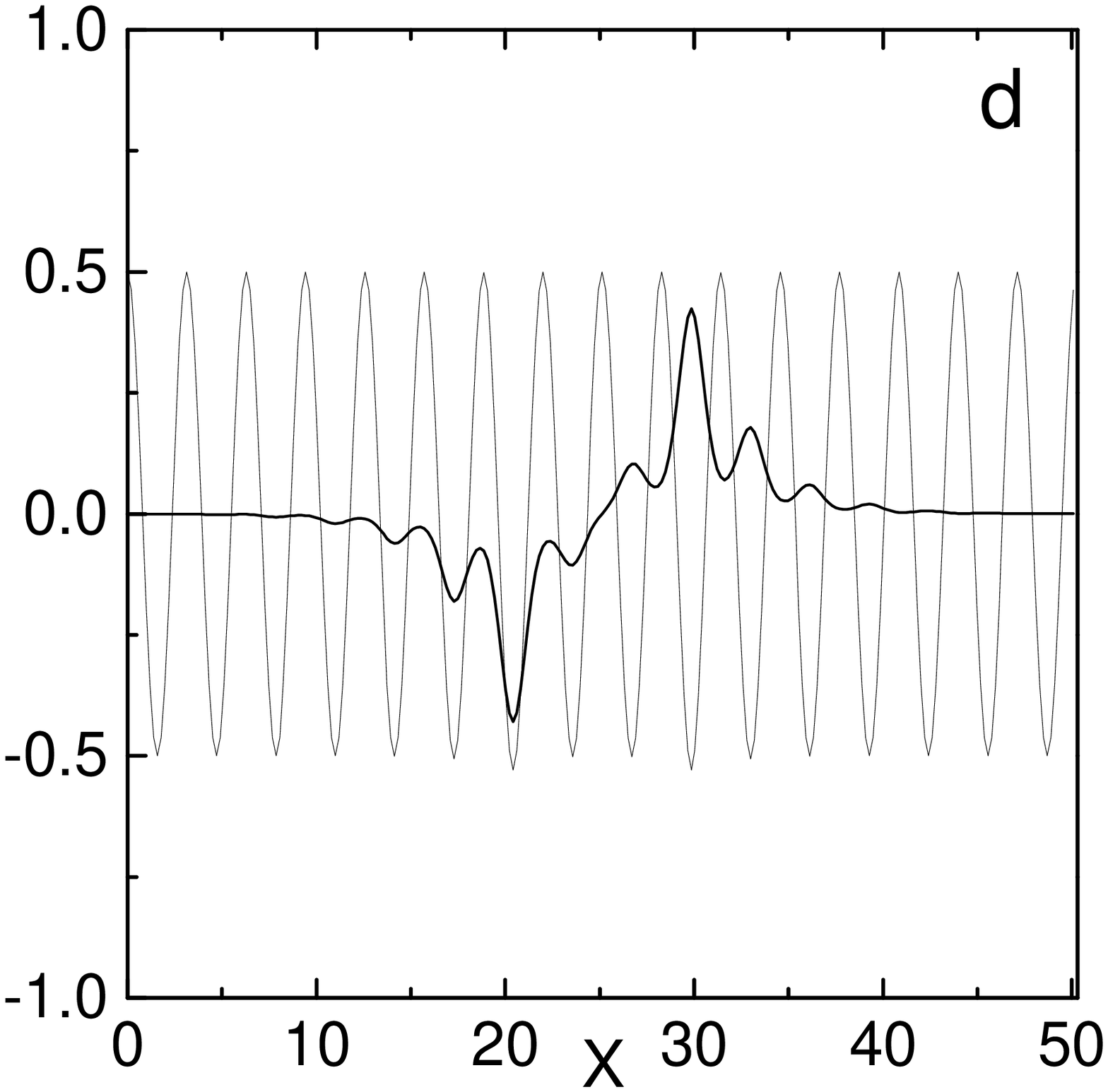}}
\centerline{
\includegraphics[width=3.8cm,height=3.cm,angle=0,clip]{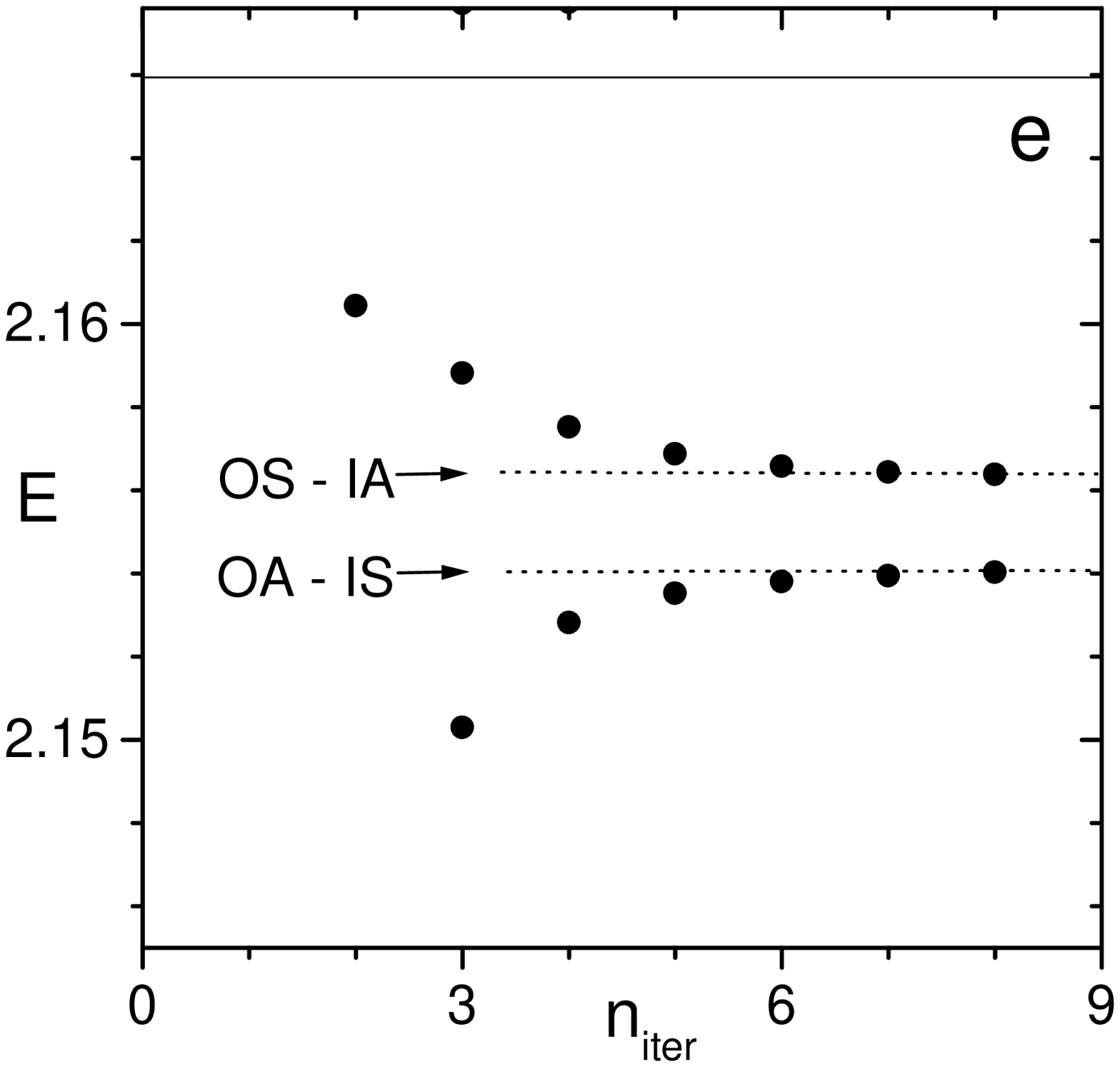}
\includegraphics[width=3.8cm,height=3.cm,angle=0,clip]{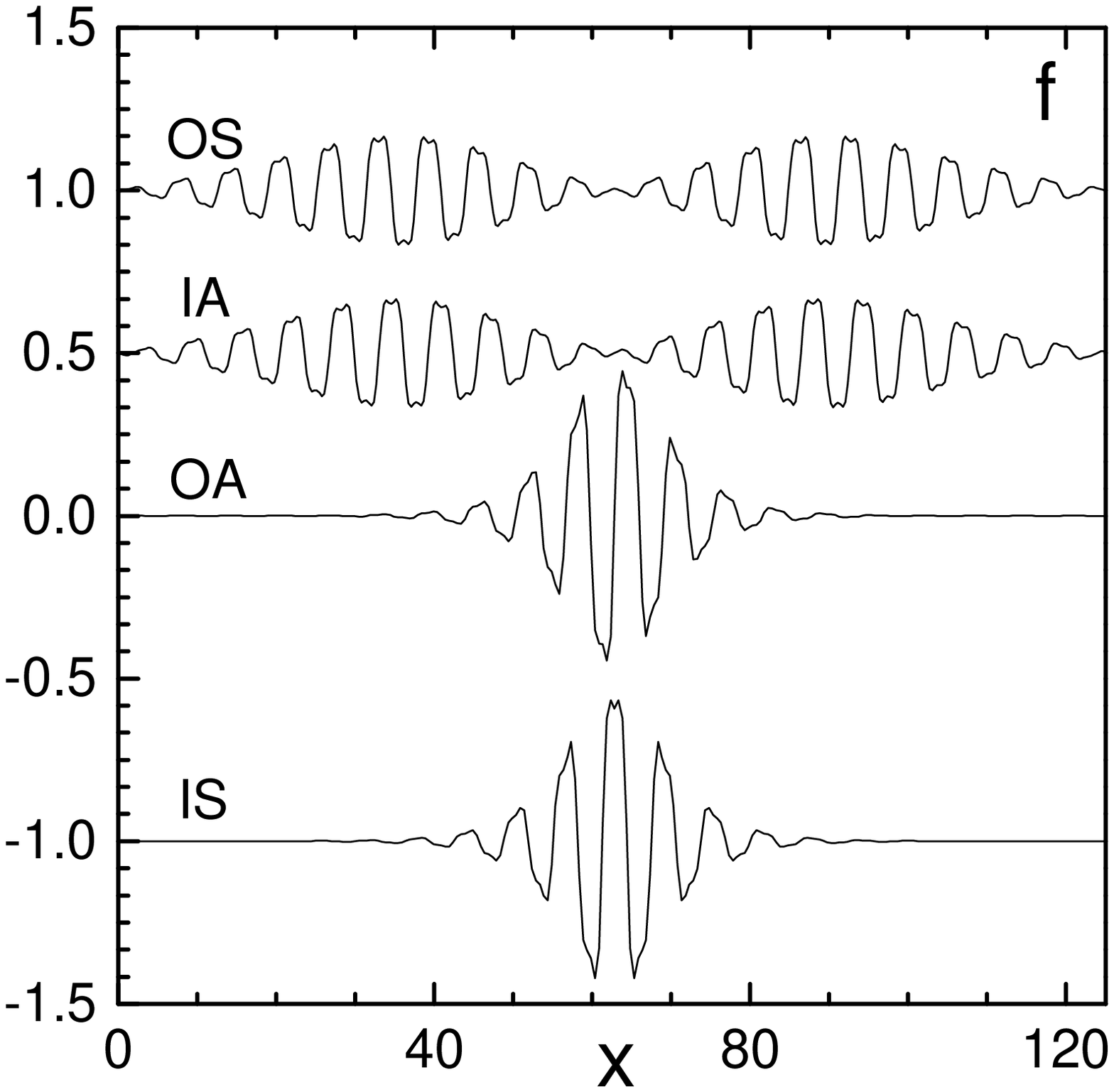}}
\caption{Wavefunctions and corresponding effective potential for
the bound states below the lowest bands of Fig 1a for attractive
interaction $\chi=-1$. Panel {\bf (a)}. OS mode and corresponding
effective potential for $E=-1.1667950$ (ground state) and $A=3$.
The effective potential was scaled by a factor 6 for graphical
convenience. Panel {\bf (b)}. Same as (a) for the IS mode at
$E=-1.0485745$ and $A=-3$. Panel {\bf (c)}. Same as (a) for the OA
mode. $E=-0.999261$. Panel {\bf (d)}. Same as (b) for the IA mode.
$E=-0.9947127$. {\bf (e)} Energy levels of the OS, IS, OA, IA,
nodes inside the gap between the first two bands. The continuous
line denotes the lower edge of the second band of Fig. 1a while
the dotted lines denote the degenerate levels. Parameters are
fixed as for corresponding modes in panels a-d. {\bf (f)}.
Wavefunctions associated to the levels in panel e. For graphical
convenience the IS mode was shifted  by -1.0 down while the IA and
OS modes were shifted up by 0.5 and 1.0, respectively.}
\label{fig2}
\end{figure}
\vskip -.2cm \noindent

For an effective construction of  these matrix elements one can
use the fast Fourier transform while for the computation of the
spectrum one can recourse to standard numerical routines for the
diagonalization of real symmetric matrices. To check the method we
consider first  the case of a linear effective potential of the
form $V_{eff}=U_{ol}=A \cos(2 x)$ for which  the eigenvalue
problem reduces to the well known Mathieu equation. In Fig. 1a we
depict the lowest part of the spectrum (notice that in this case
there is no SC procedure due to the linearity of the problem) from
which we see the appearance of a band structure with band edges
which exactly coincide with the values obtained for the Mathieu
equation (for high energy bands to get good accuracy one needs
to increase $N$). In this paper we are mainly interested in the
localized states associated with the lowest two bands (i.e., the
ones physically most relevant), and for this purpose the choice of
$N=256$ will be adequate for most of the calculations.
\begin{figure}
\centerline{
\includegraphics[width=4.cm,height=3.6cm,angle=0,clip]{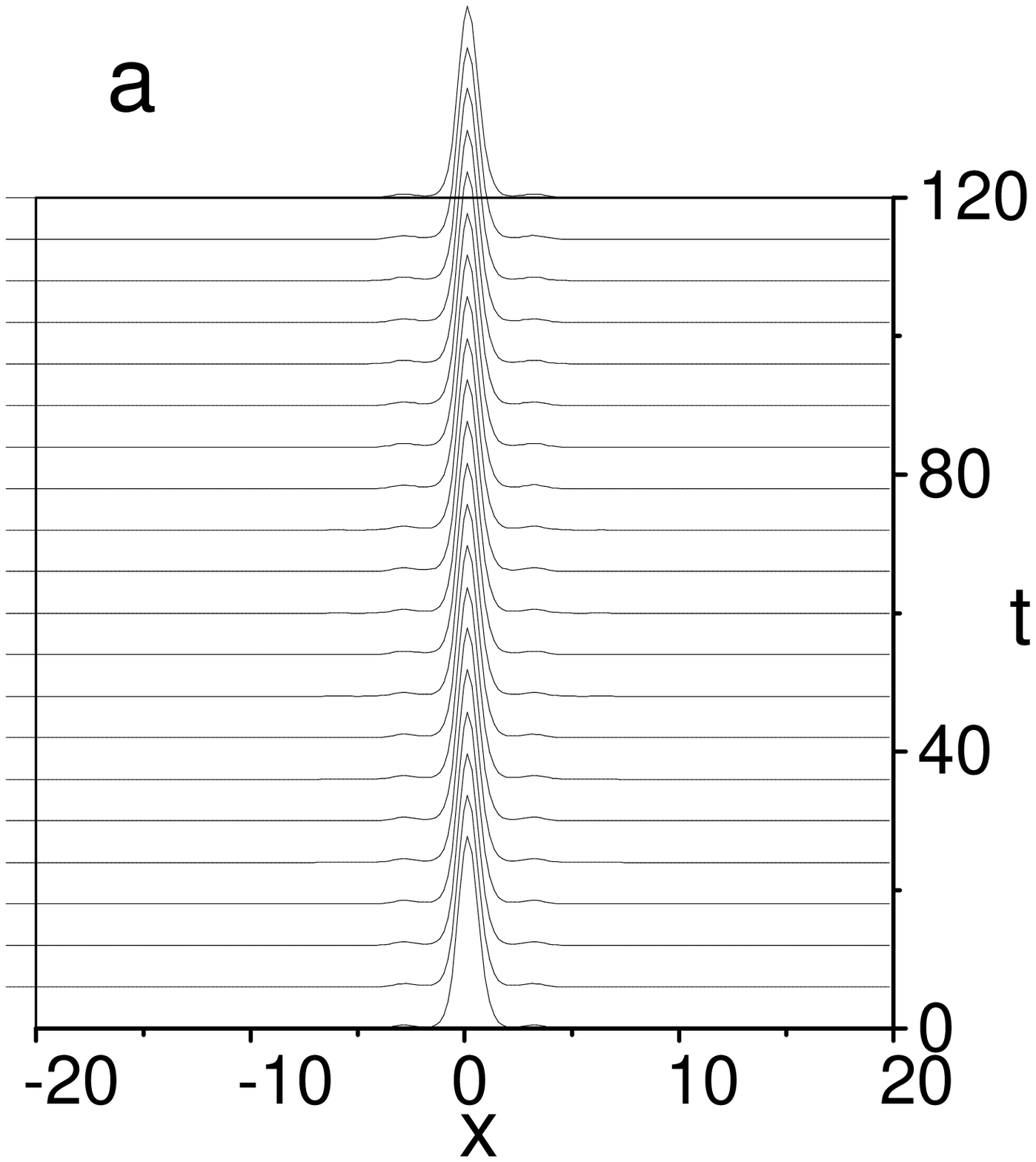}
\includegraphics[width=4.cm,height=3.6cm,angle=0,clip]{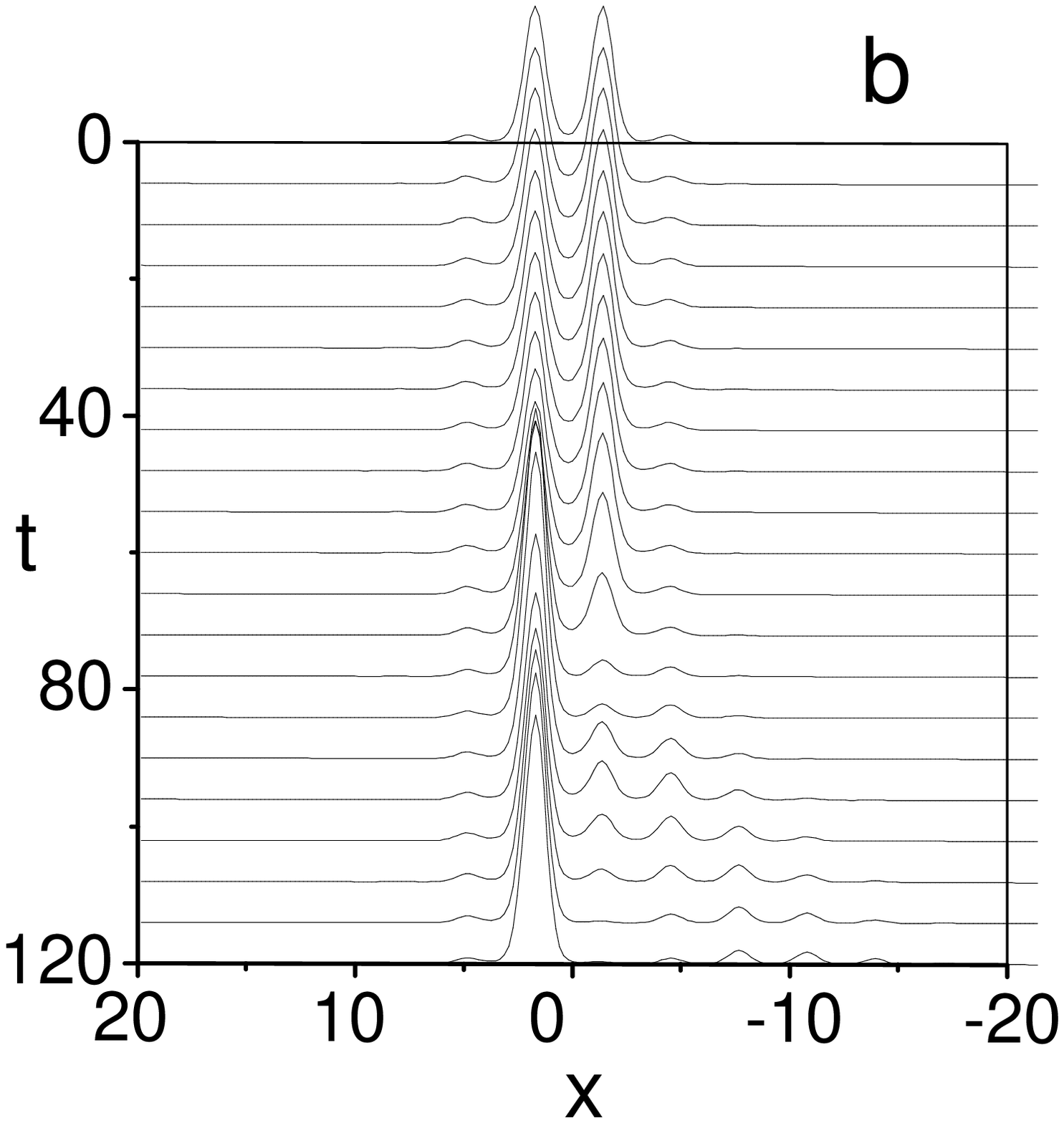}}
\caption{Panel {\bf (a)}. Time evolution of the OS bound state of
Fig 2a as obtained from GPE. Panel {\bf (b)}. Same as in panel (a)
for the IS mode of Fig. 2c.} \label{fig3}
\end{figure}
\begin{figure}
\centerline{
\includegraphics[width=3.8cm,height=3.4cm,angle=0,clip]{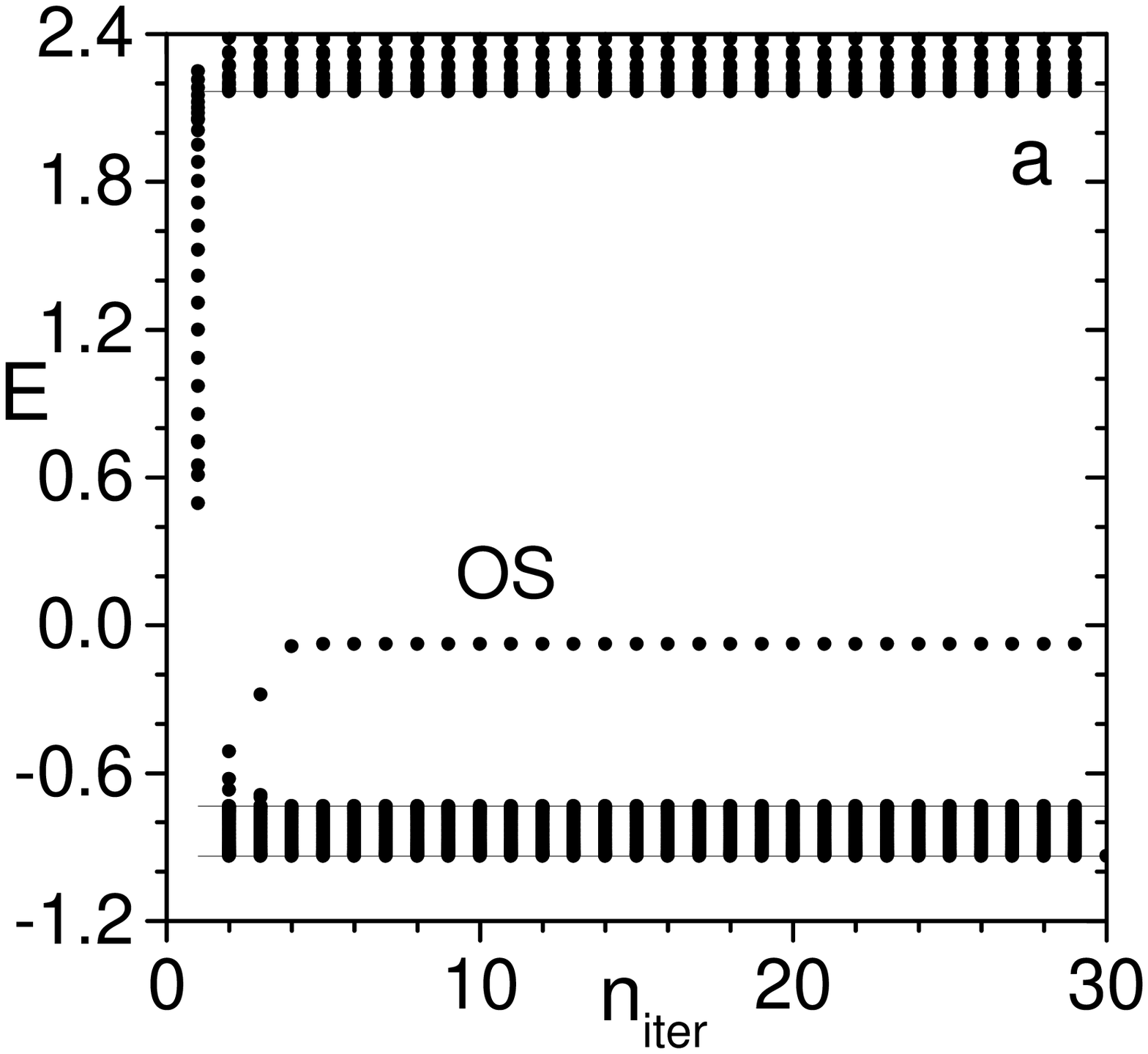}
\includegraphics[width=3.8cm,height=3.4cm,angle=0,clip]{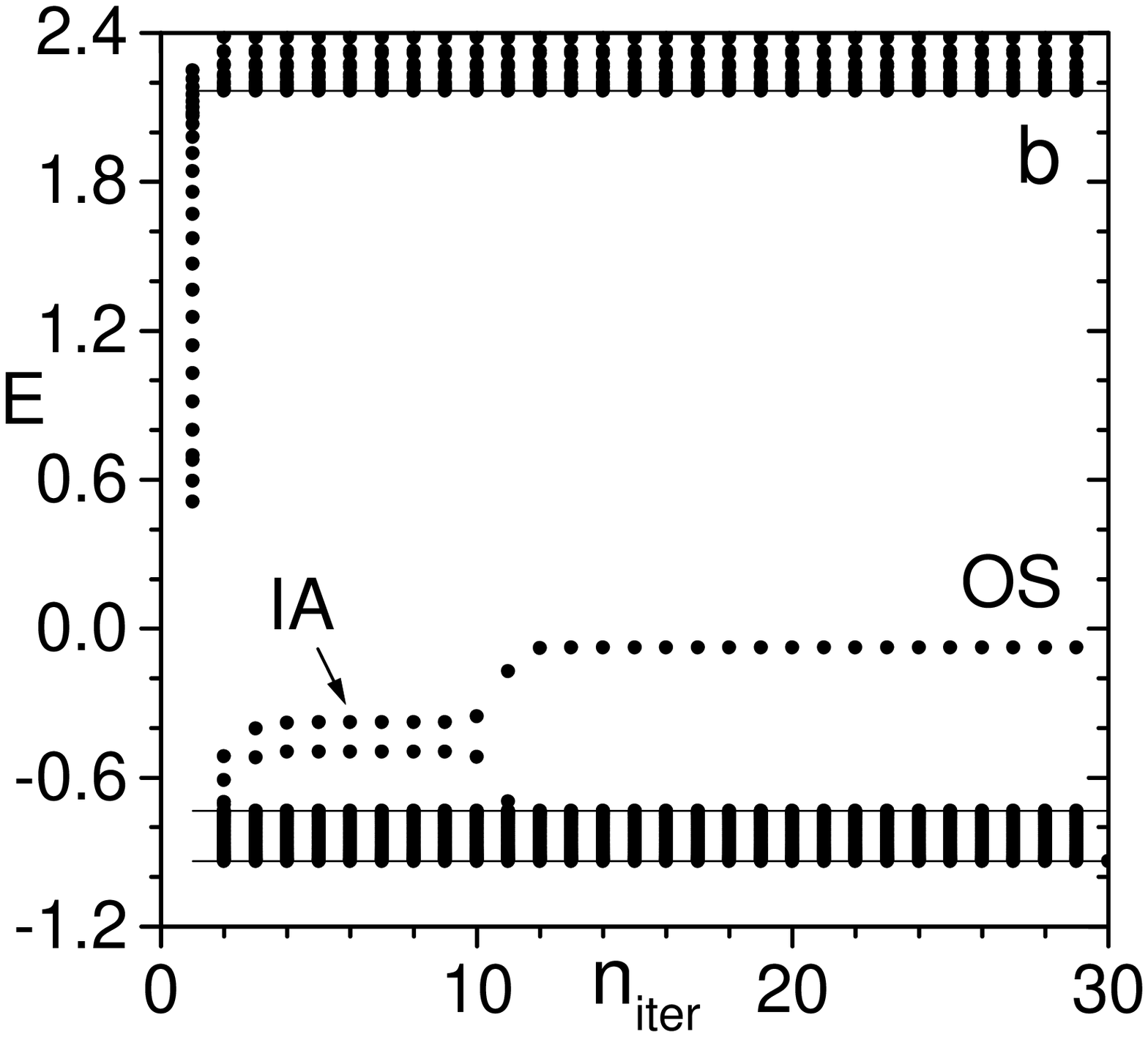}}
\caption{Energy spectrum in correspondence of the localized states
above the lowest band of Fig. 1a for repulsive interactions
$\chi=1$. Panel {\bf (a)}. Spectrum associated to the OS mode.
Parameters are $A=-3, N=256, L=40 \pi$. Panel {\bf (b)}. Same as
panel (a) but for the IA mode with  $A=3$. The continuous lines
denote exact band edges of the Mathieu equation. } \label{fig4}
\end{figure}
%
%\vskip -2.6cm
In Fig. 1b we show how the lowest band of panel 1a is modified by
the nonlinear potential $V_{eff}(x)=A \cos(2 x)+ \chi |\psi_0|^2$,
where $\psi_0$ is taken to be the ground state of the system,  for
the case $\chi<0$ (negative scattering length). A bound state
below the band which rapidly converges to a constant value is
quite evident. One can see that the state forms from the lower
edge of the band and is accompanied by a rearrangement of the
extended states inside the band.  The corresponding eigenvector is
depicted in Fig.2a together with its effective potential. Notice
that the potential has an attractive character (potential well)
and the bound state is symmetric around a minima of the OL, i.e.,
it resembles the onsite-symmetric intrinsic localized mode (ILM)
of nonlinear lattices (NL) \cite{sievers}. In the following we
shall call it the onsite symmetric (OS) bound state. By shifting
the phase of the OL by $\pi$ (i.e. by changing the sign of $A$)
one obtains an eigenstate centered on maxima instead than on minima.
This bound state is depicted in Fig. 2b and in analogy with NL we
shall call it the intersite symmetric (IS) mode. The corresponding
spectrum is reported in Fig. 1c. Notice that the IS mode
corresponds to the plateau formed just before the decay into the
OS mode occurs as shown in panel 1d (also note the appearance of
an intersite-symmetric (IA) excited level in Fig. 1c which is
absorbed into the band in correspondence of the IS-OS decay).
\begin{figure}\centerline{
\includegraphics[width=3.8cm,height=3.cm,angle=0,clip]{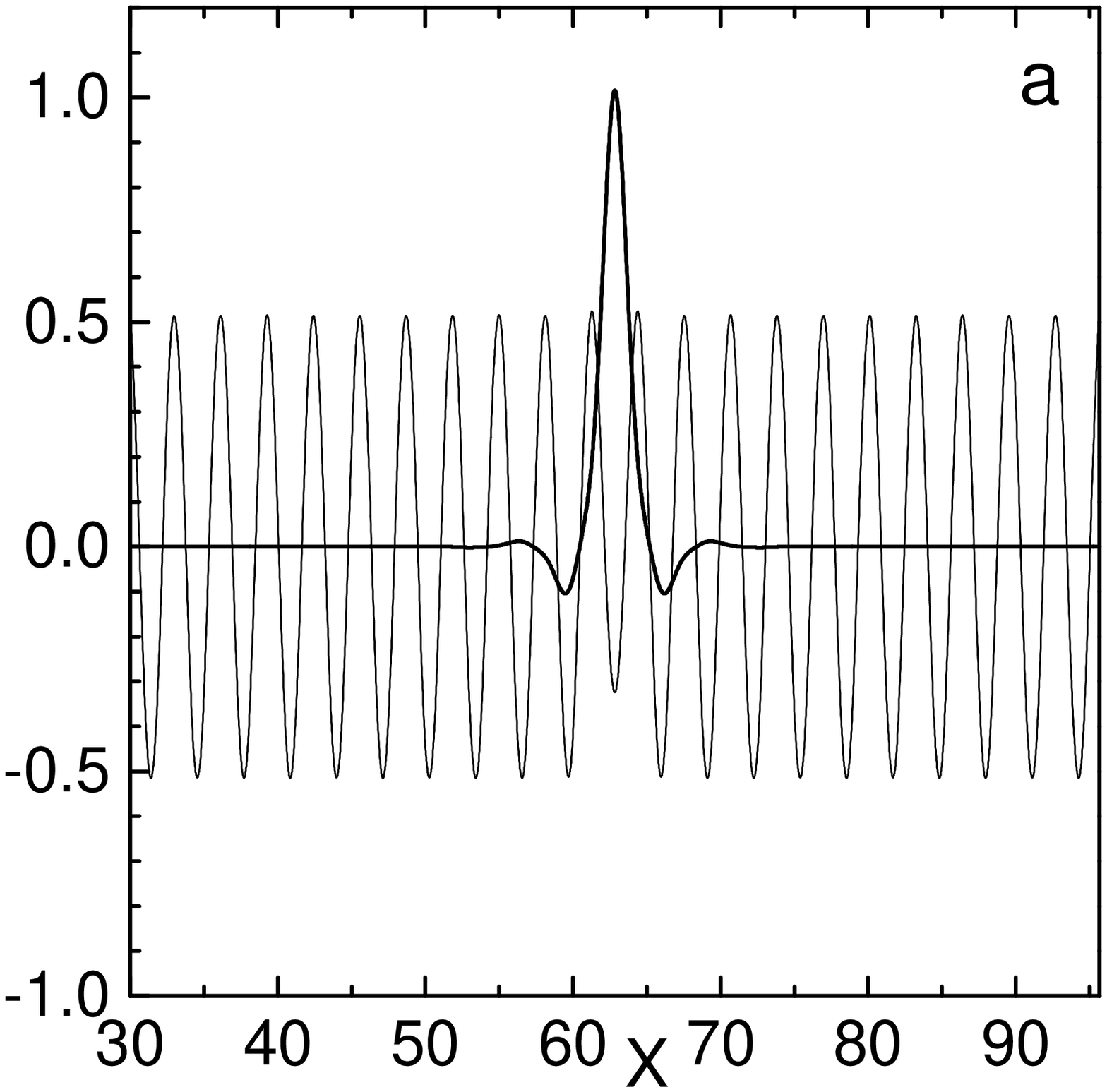}
\includegraphics[width=3.8cm,height=3.cm,angle=0,clip]{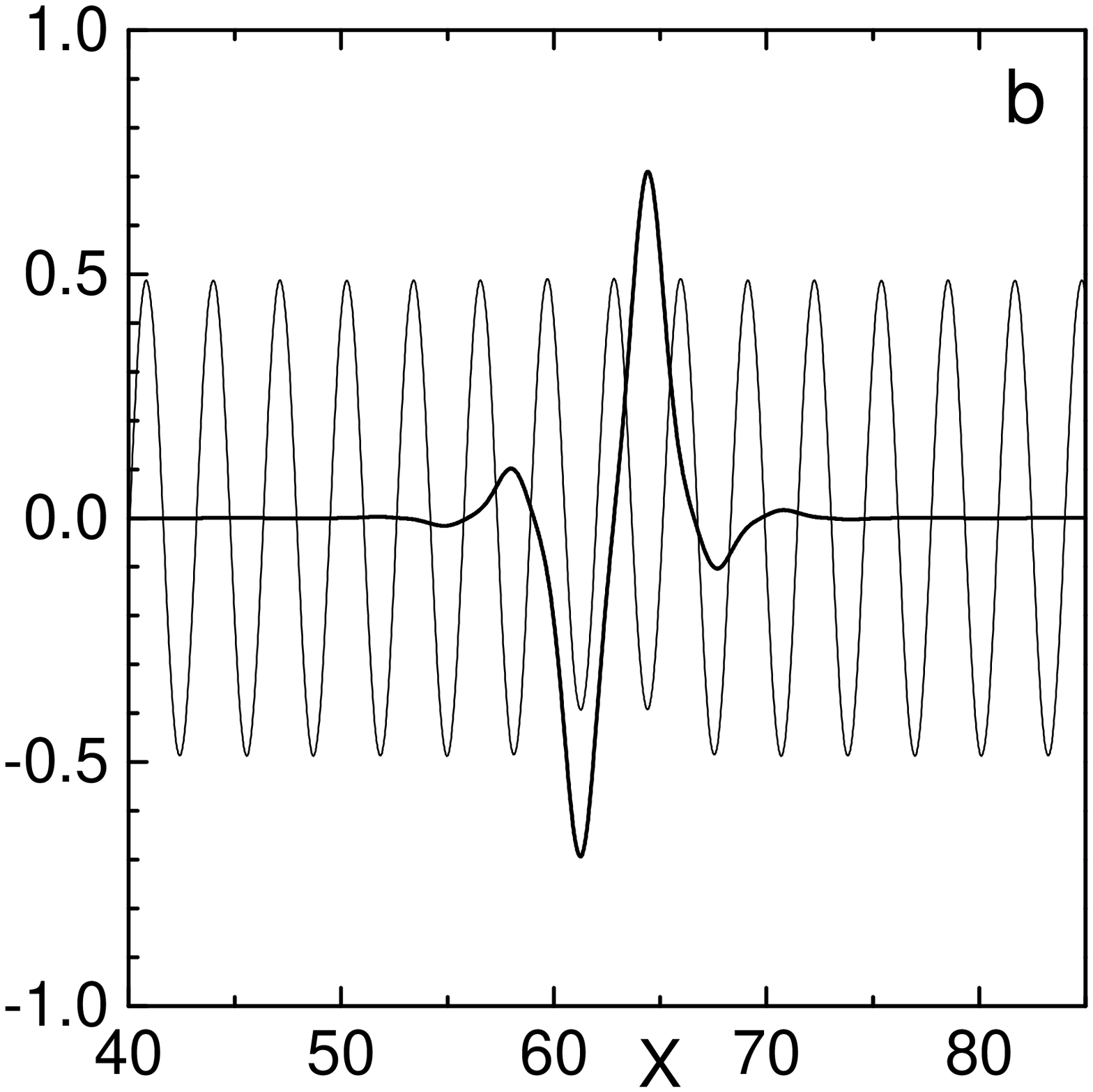}}
\centerline{
\includegraphics[width=3.8cm,height=3.cm,angle=0,clip]{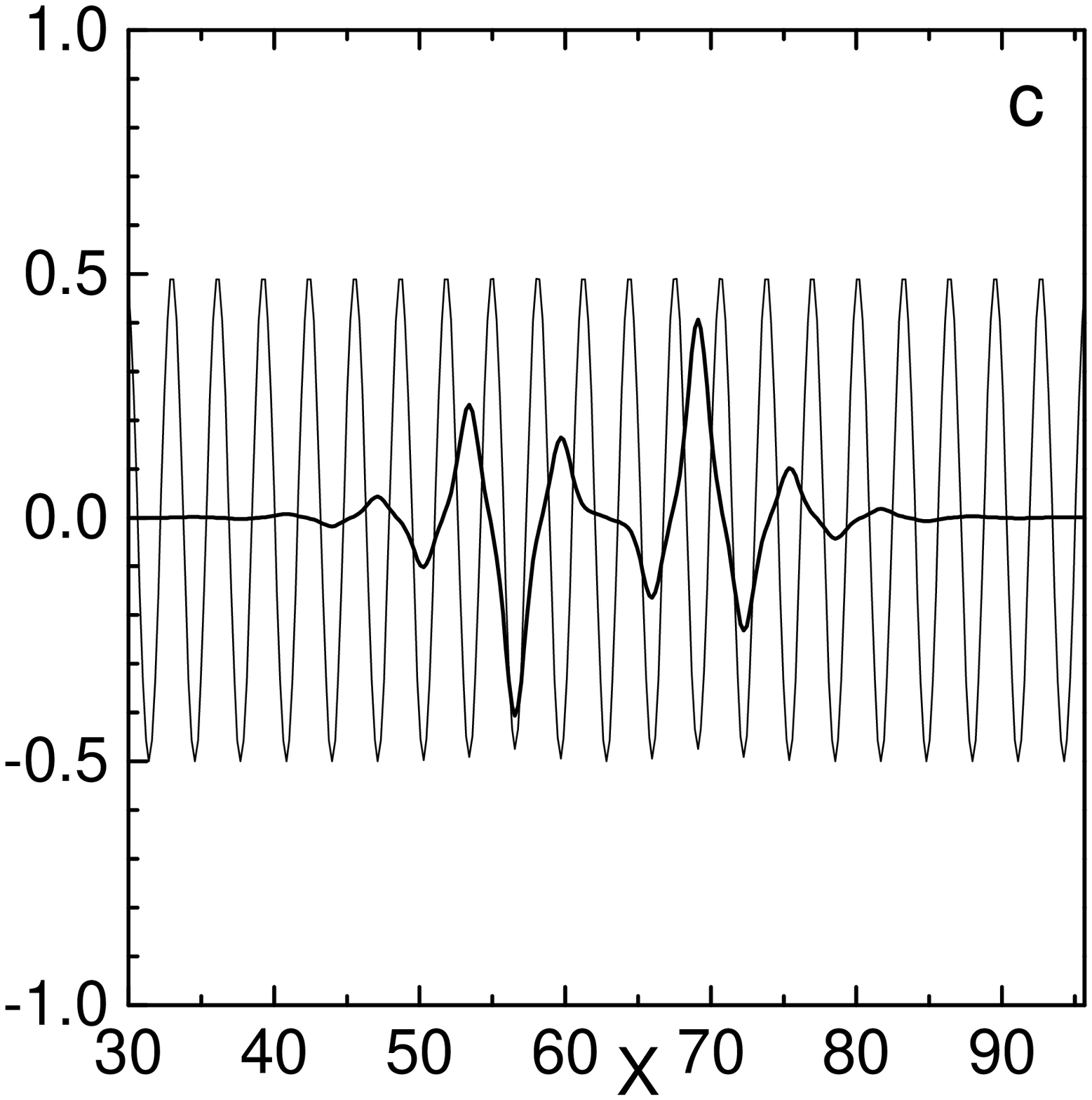}
\includegraphics[width=3.8cm,height=3.cm,angle=0,clip]{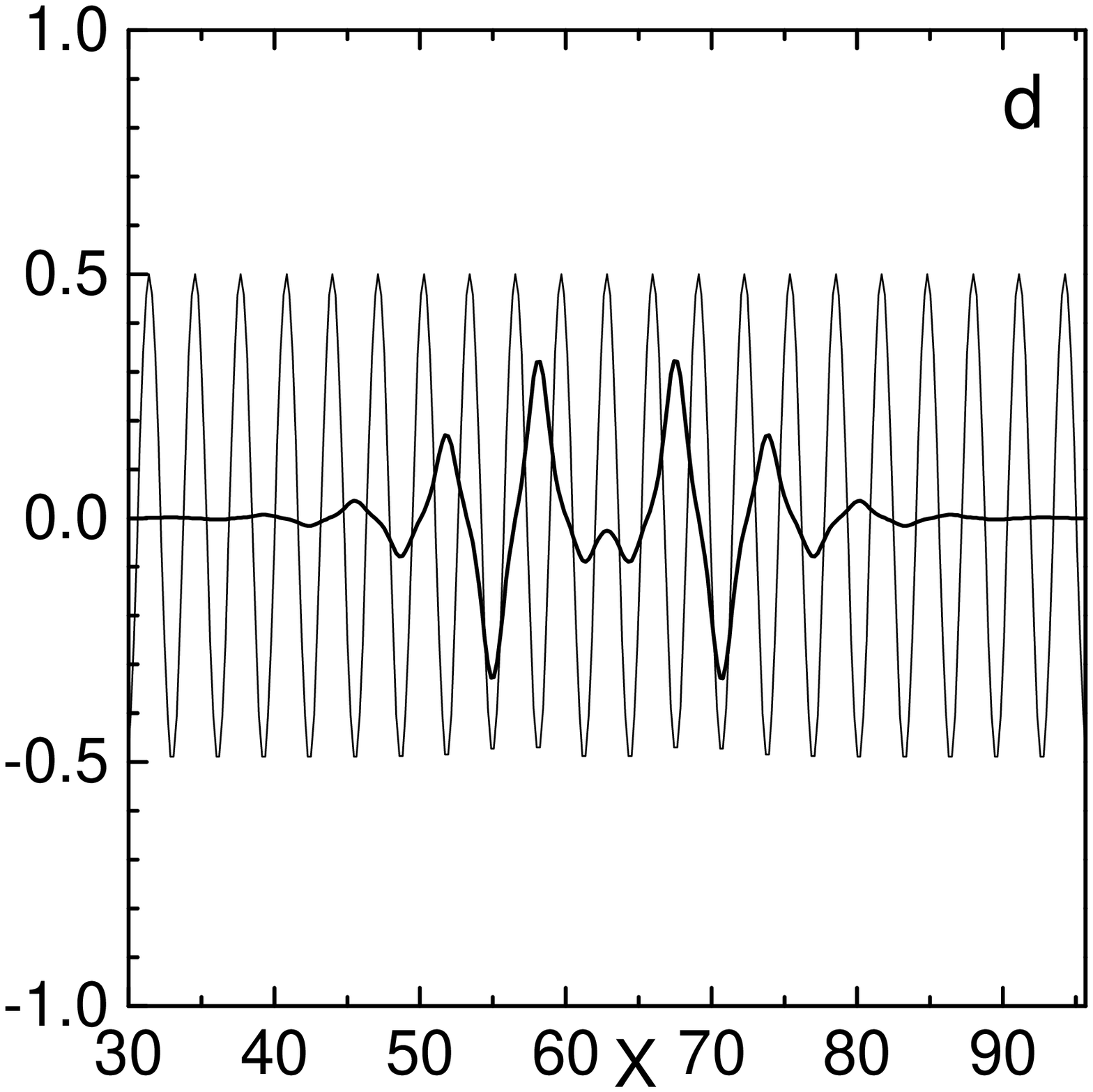}}
\caption{Wavefunctions and effective potentials of the bound
states levels of Fig. 4 a,b, for the repulsive case $\chi=1$.
Panel {\bf (a)}. OS mode with corresponding effective potential
(thin line). Energy is $E=-0.078355$ and $A=-3$. Panel {\bf (b)}.
Same as in Panel (a) for the IA mode. $E=-0.376645$, $A=3$. Panel
{\bf (c)}. Same as in panel (a) for the OA mode. $E=-0.683070$.
Panel {\bf (d)}. Same in panel (b) for the IS mode. $E=-0.691676$.
Parameters are fixed as $N=256$, $L=40 \pi$ for panels (a), (b),
and $N=512$, $L=40 \pi$ for panels (c), (d). The effective
potentials have been reduced by a factor $6$ for graphical
convenience. } \label{fig5}
\end{figure}
We have checked that these bound states coincide with the ones
obtained with the approach of Ref. \cite{alfimov} for the same
values of energy. The stability of the OS mode and the decay of
the IS mode into the OS state was checked by direct numerical
integrations of the GPE (see Fig.3). To obtain the onsite
asymmetric (OA) mode one needs to take the first excited state
$\psi_1$ as effective potential in the SC procedure. This indeed
produces an exact soliton solution of the GPE of type OA as shown
in Fig.2c. A shifting of the potential by $\pi$ produce the
intersite asymmetric (IA) mode of Fig. 2d. These solutions have
the same energies and are more unstable than the IS mode (they,
however, do not decay into the ground state but get mixed with the
extended states in the band).
In general, the effective potentials can be taken of the form
$\hat V_{eff}=\hat V_{ol}+ \chi |\hat \psi_n (x)|$ with $\psi_n$
the n-th eigenstate of the eigenvalue problem in (\ref{schro}). If
the energy  of $\psi_n$ lies outside the bands a localized mode of
the type described above is produced, while if it lies inside a
band, extended states which are nonlinear analogue of the Bloch
states \cite{pethick}, are produced. From this we conclude that
both localized and extended solutions of the GPE are exact quantum
eigenstates of the Schr\"odinger equation with a suitable
effective potential.
%Notice that for $\chi<0$, the
%nonlinear part of the effective potential lower the energy, so
%that the linear states at the bottom of the bands are pulled down
%in the forbidden zones as $|\chi|$ is increased, as one can see
%from Fig.1b,c.
From the above analysis one expects that below each higher energy
bands, eigenstates of the same symmetry type as the ones found for
the lowest band should exist. This is precisely what we show in
Fig.s 2e, 2f for the energy spectrum and the corresponding
eigenstated  in the gap below the second band.
\begin{figure}\centerline{
\includegraphics[width=3.8cm,height=3.6cm,angle=0,clip]{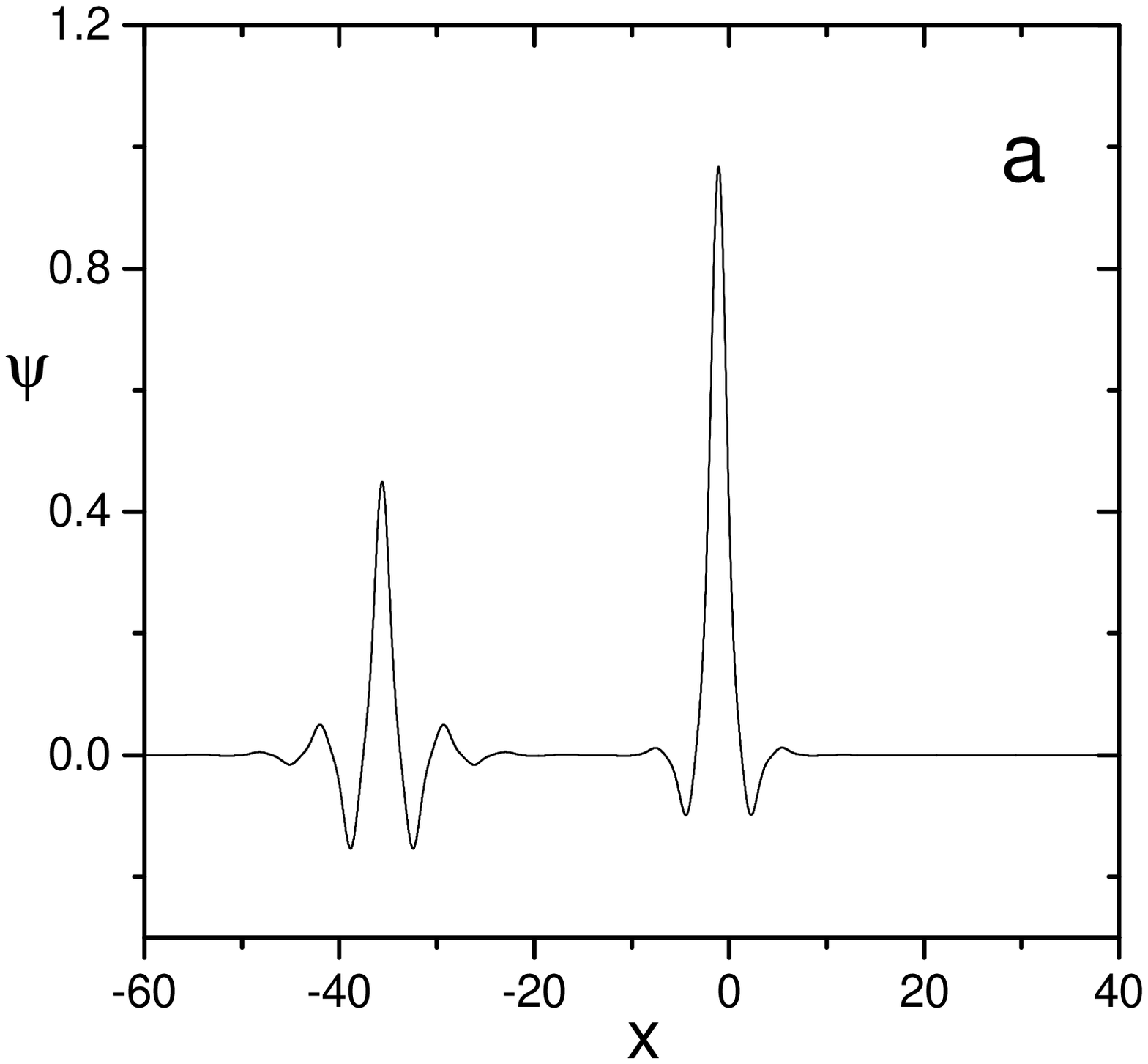}
\includegraphics[width=3.8cm,height=4.6cm,angle=0,clip]{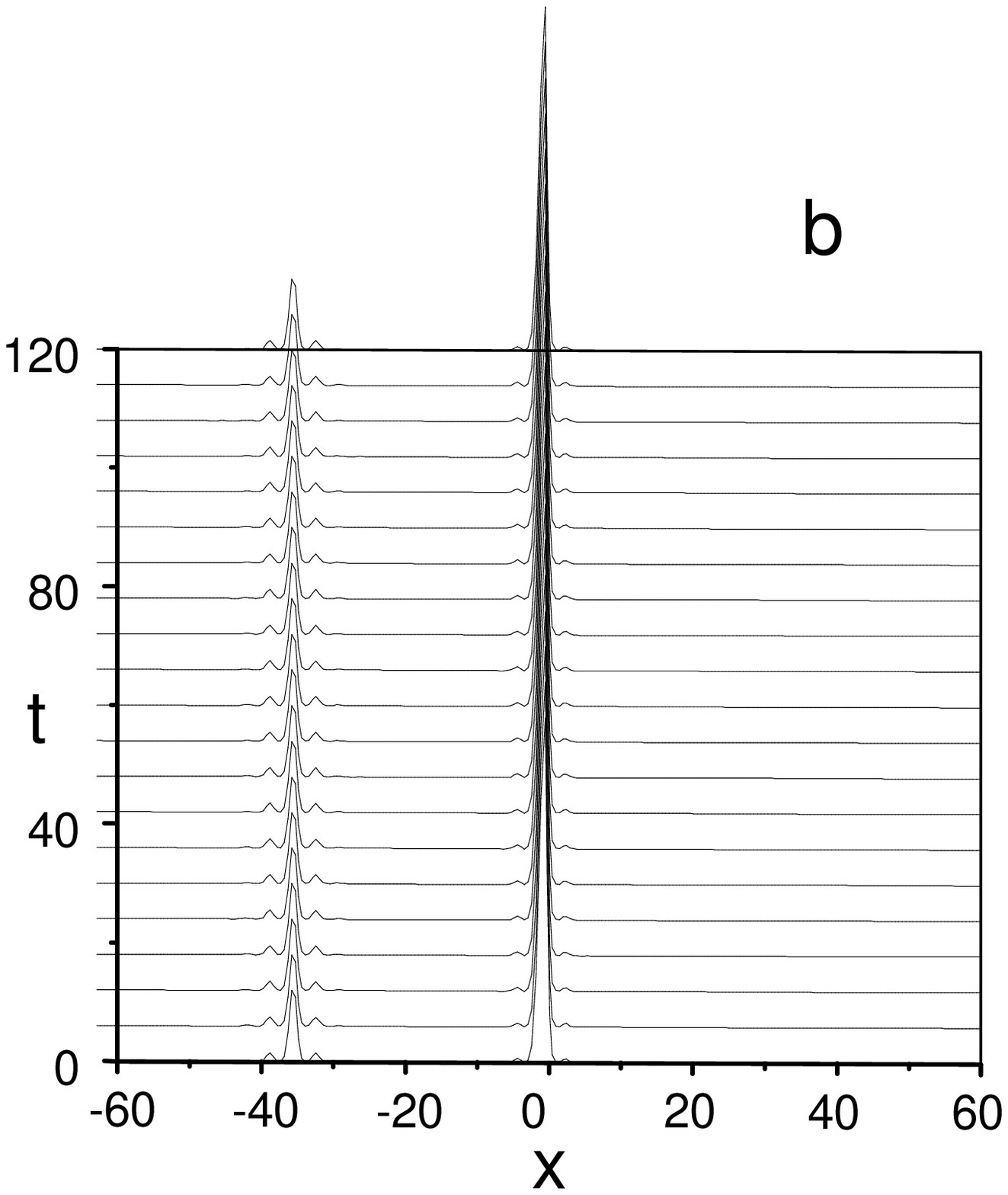}}
\caption{Panel {\bf (a)}. Two soliton bound state of the repulsive
GPE obtained from the SC procedure by using as effective potential
$V_{eff}=V_{ol}+|\frac 12 \psi_1-\psi_2]^2$ where $\psi_1, \psi_2$
denote two eigenfunctions at the top of the first band. Panel {\bf
(b)}. Time evolution (modulo square) of the bound state in panel
(a) taken as initial condition for the integration of the full
GPE. } \label{fig6}
\end{figure}
Notice that the OA and the IS eigenstates are degenerated (the
same occurs also to the OS and IA modes). The OA and IS bound
states are both very stable while the energy levels of the OA and
IS modes, after establishing a plateau similar to the one in Fig
1c, become unstable (the energy oscillates between this level and
the lower edge of the second band). The instability of these modes
can be understood as an hybridization  of the state (being very
close to the band edge) with extended states of the second band
and is confirmed by direct numerical integration of the PDE
system.

Similar localized modes can be found also for repulsive
interactions ($\chi>0$), the main difference with the previous
case being that now the states appear in the gap above the band
edges instead than below.
%(this is because the nonlinearity by increasing the energy
%pulls the linear states in the upper part of the bands in the
%gaps).
This is shown in Fig. 4 for the lowest energy levels inside the
first gap. The corresponding eigenvectors are shown in Fig.5
together with their effective potentials. Notice that the
potential has a local repulsive character (it increases in
correspondence of the states) so that these bound states could not
exist without the OL. We remark that localized modes similar to
the ones described in this paper were found also in
atomic-molecular BEC using an approach based on Wannier functions
\cite{abdullaev}.
%One can think to
%this state in analogy with the case of Bloch electrons in a
%crystal, as an bound state of holes having negative effective mass
%(the modes for $\chi<0$ would then corresponds to bound states of
%electrons with positive effective mass).
%In summary, we see that the  above SC approach has the following
%advantages: i)allows to establish a link between gap solitons and
%quantum bound states of a linear Schr\"oedinger problem, ii)
%allows to get localized and extended states, as well as
%information on their stability: iii) converges quite  rapidly; iv)
%can be applicable to arbitrary potentials in arbitrary dimensions
%(in these cases, however, the diagonalization problem becomes
%computationally more expensive (it grows as $N_p^{2d}$)).

Besides localized and extended states, the SC procedure allows
to construct full nonlinear bands in reciprocal space (we omit
details for brevity). It is worth remarking that more complicated
set of solutions of the GPE can be constructed with the SC
procedure by taking as effective potentials linear combinations of
eigenstates. An example of this is shown in Fig.6 for the case of
repulsive interaction. We see that a linear combination of two
eigenstates leads to a bound state with two humps which
corresponds to a multi-soliton solution of the GPE (see panel b).
This is a general property and  we  conjecture that {\it all
solutions of the periodic GPE (and more in general of the NLS-like
equations with arbitrary potentials) can be obtained with the SC
method taking all possible combinations of linear eigenstates as
effective potentials}.

Before closing this paper we wish to remark that the above bound
state interpretation has important consequences on the
delocalizing transition \cite{flach} of localized solutions of the
GPE in OL. Since in a 1D potential bound state always exists,  it
follows from the above analysis that no delocalizing transition of
a BEC soliton can occur in this case. On the contrary, for $D \geq
2$ a finite potential depth is required to form a bound state,
this implying that a soliton delocalization transition can occur.
A detailed investigation of the delocalizing transition of BEC
solitons in OL will be reported elsewhere \cite{bs03}.

It is a pleasure to thank Prof. S. De Filippo and Dr. B.
Baizakov for interesting discussions. Financial support from a
MURST-PRIN-2003 Initiative, and from the European grant LOCNET no.
HPRN-CT-1999-00163, is also acknowledged.

\end{multicols}

\end{document}